\documentclass[
reprint,
 amsmath,amssymb,aps,pre
]{revtex4}

\usepackage[            
pdftex,                 
pdfborder= 0 0 0,
colorlinks=true,        
citecolor=green,        
linkcolor=cyan,         
menucolor=blue,         
urlcolor=magenta,       
bookmarksopen=true
]{hyperref}

\usepackage{graphicx}
\usepackage{dcolumn}
\usepackage{bm}

\usepackage[labelfont=bf,labelsep=period,justification=raggedright]{caption}

\makeatletter
\renewcommand{\@biblabel}[1]{\quad#1.}
\makeatother

\date{}

\newcommand{\z}{Z\kern-0.6emZ}

\newcommand{\be}{\begin{equation}}
\newcommand{\ee}{\end{equation}}
\newcommand{\E}{I\kern-0.3emE}

\usepackage{bm}

\begin{document} 


\title{Phase transition from egalitarian to hierarchical societies driven by  competition between cognitive and social constraints} 
\author{Nestor Caticha}\email{nestor@if.usp.br}
\author{Rafael Calsaverini}\email{rafael.calsaverini@gmail.com}
\affiliation{Dept. de F{\'\i}sica Geral, Instituto de F{\'\i}sica, \\
Universidade de S\~ao Paulo, 05508-090, S\~ao Paulo-SP, Brazil}

\author{Renato Vicente}
 \email{rvicente@usp.br}
\affiliation{Dept. Matem\'atica Aplicada, Instituto de Matem\'atica e Estat{\'\i}stica, 
Universidade de S\~ao Paulo, 05508-090, S\~ao Paulo-SP, Brazil}
\date{\today}
\begin{abstract}
Empirical evidence suggests that social structure may have changed from hierarchical to egalitarian and back
along the evolutionary line of humans. We model a society subject to competing 
cognitive and social navigation constraints. The theory predicts that the degree of
hierarchy decreases with encephalization and increases with group size.
Hence hominin groups may have been driven from a phase with hierarchical order  to a phase with egalitarian structures
by the encephalization during the last two million years, and back to
hierarchical due to fast demographical changes during the Neolithic. The dynamics in 
the perceived social network shows evidence in the egalitarian phase of the observed phenomenon of Reverse Dominance. 
The theory also predicts for modern hunter-gatherers in mild climates a trend towards an 
intermediate hierarchy degree and a phase transition for harder ecological conditions. In harsher climates societies  would tend to be
more egalitarian if organized in small groups but more hierarchical
if in large groups. The theoretical model permits organizing the available data 
in the cross-cultural record (Ethnographic Atlas, N=248 cultures) where the symmetry breaking 
transition can be clearly seen.
\end{abstract}
\pacs{Valid PACS appear here}
\maketitle

\section{Introduction} 

Behavioral phylogenetics makes it  plausible that the common ancestor of {\it Homo} and {\it Pan} genera
had a hierarchical social structure \cite{Knauft1991,Boehm1999,Boehm12,Dubreuil,Dubreuilbook}.
Paleolithic humans with a foraging lifestyle, however, most likely had a largely egalitarian society and yet
hierarchical structures became again common in the Neolithic period.
Contemporary illiterate societies 
fill the ethological spectrum  \cite{Vehrencamp1983}
from egalitarian  to authoritarian and despotic. 
This non-monotonic journey, a so called U-shaped trajectory, along the 
egalitarian-hierarchical spectrum  during human evolution, 
 was stressed by Knauft \cite{Knauft1991} and has defied theory despite several attempts of anthropological explanation
\cite{Knauft1991}. Our approach
to the study of social organization
uses tools of information theory and statistical mechanics. It is inspired in previous work by 
Terano {\it et al.} \cite{terano2008, terano2009} on a different problem, the emergence of money in a barter society
as a consequence of limited cognitive capacity. 
We model the perception by each agent of the social network
of its society, taking into account  cognitive constraints 
and social navigation demands, which define the informational constraints adequate to 
a probabilistic description.
According to the model, whether an egalitarian-symmetric or
hierarchical-broken symmetry  state occurs depends on a scaling
parameter which grows with
cognitive capacity and decreases with group size, 
modulated by  a Lagrange multiplier which can be interpreted as an environmental pressure. 
The hypothesis that social perceptions mediate motivations
and hence  possible behaviors (PMB hypothesis), permits making  predictions 
about the effect of these variables in the 
probable forms of social organization. Since there was a massive increase in encephalic mass in the last 
two million years, our theory expects a phase transition towards a more
egalitarian  social organization to occur. As food producing and storage methods permitted
the populational increase in the Neolithic, the scaling parameter decrease permitted 
a reversal to hierarchical structures.
 
Furthermore, the same model makes
predictions about a totally different  empirical situation, dealing with  
the influence of ecology on the expected hierarchy
of modern human groups. The theory suggests a form of looking at 
the available ethnographic data  \cite{ethnoatlas, Murdock2006} and
allows a new interpretation of observed patterns involving social structure, community size and environment  in terms
 of a  competition between cognitive constraints and social
navigation demands and a symmetry breaking phase transition. The bifurcation
suggested by the theory is seen in the ethnographic records. The empirical data can be found  in
the Murdock's Ethnographic Atlas \cite{ethnoatlas}, which resulted from the {\it tour de force} attempt to compile all ethnographic available knowledge that can be represented in a quantitative form. 

We are studying the properties of finite size groups and so cannot take
the thermodynamic limit. The changes in behavior are therefore not singular 
but we still find that the language of phase transitions is adequate, after all 
from a technical point of view the infinite size limit is a tool to simplify 
the mathematical treatment of very large systems.

\section{The theory and the model \label{teoria}}
In a group of $n$ agents, each agent of the group will have a perceived
social web of interactions 
 represented by a graph. An important distinction has to  be made between 
the social network, that say an ethnographer might describe and the perceived
social network of a particular agent.
Each  vertex of a graph stands for a represented agent of the group.
In this representation of the social web, undirected edges joining any pair
of vertices might be present or not. 
An edge links two represented agents
 if their social relation is known
by the owner of the graph. Since inference depends on the available
information, these graphs might differ from agent to agent. 
Call ${\bf S}^i$ the representation of the social web by agent $i$, given by
a set $\{ s^i_{jk}\}$
of variables that can take values either zero or one. The indices 
$i, j$ and $k$ run from $1$ to 
 $n$ and every $s^i_{jk}$ is symmetric 
with respect to interchange of the lower indices. 
If   $ s^i_{jk}=1$ then agent $i$
has knowledge of the social relation, the capacity to cooperate and
form coalitions or the antagonism between agents $j$ and $k$, 
while  if $ s^i_{jk}=0$ this relation is unknown to agent $i$. 
Known alliances, feuds or neutral interactions 
are represented by  $ s^i_{jk}=1$.
The total number of memorized social relations is
\begin{equation}
N^i_{cog} = \sum_{j, k=1}^n s^i_{jk}/2
\end{equation} 
the number of edges in the social 
web representation of agent $i$. This is a cognitive contribution to
the cost of a given representation ${\bf S}^i$.

\begin{figure}
\center
\includegraphics[width=0.7\textwidth]{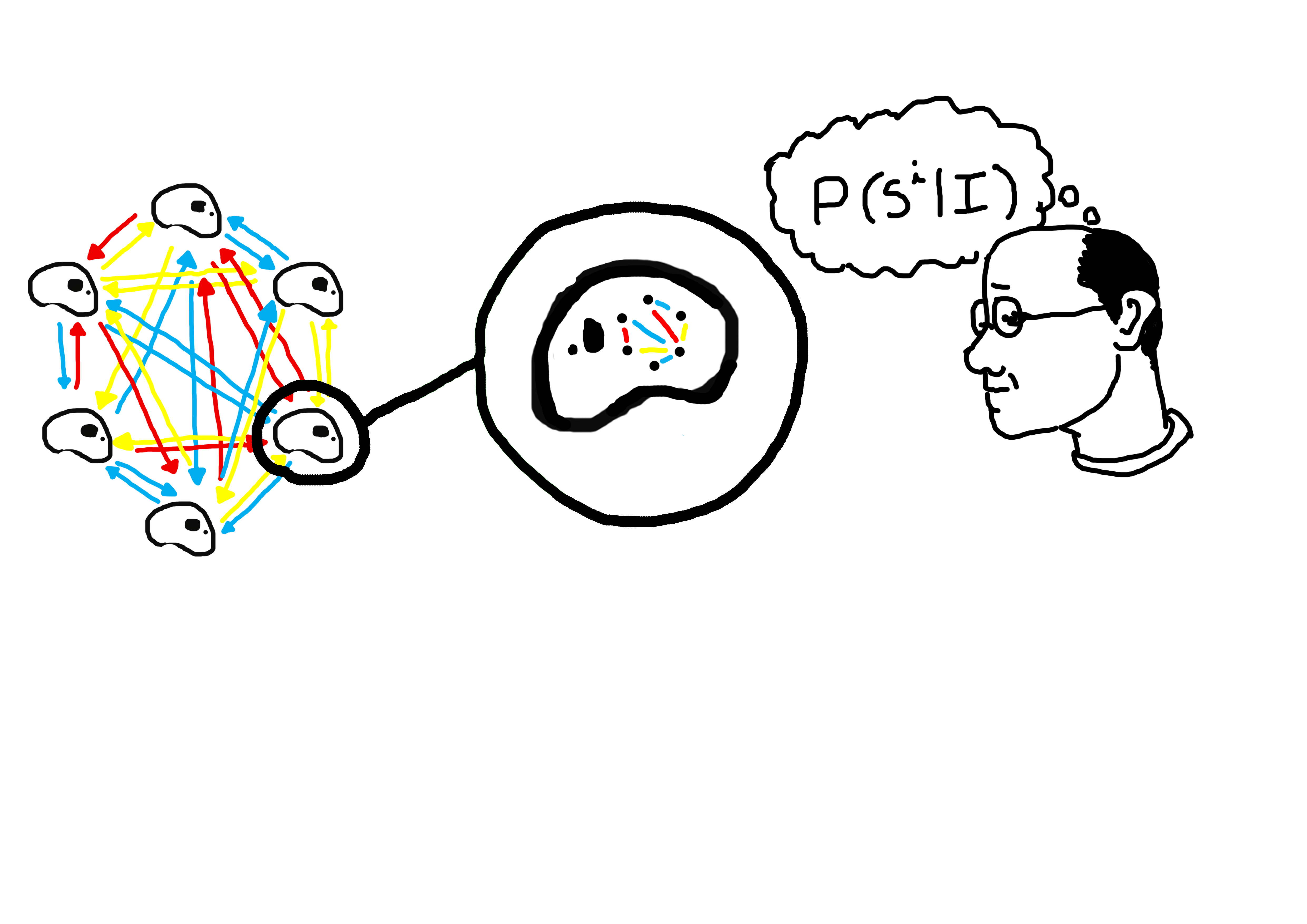}
\vspace{-1.5cm}
\caption{Each agent has a perceived social web of interactions 
 represented by a graph. Each vertex of a graph stands for a represented agent of the group. An edge in this perceived web links two agents if their social relation is known by the owner of the graph. Using the methods of 
information theory entropic inference, we attribute a probability $P( {\bf S}^i |I)$ that agent $i$ perceives a network $S^i$, conditional on any   
available information $I$.}
\label{socialcog}
\end{figure}

Now consider the agent $i$'s social cost 
for not knowing a given social relationship, when  $ s^i_{jk}=0$.  
For two  agents $j$ and $k$ there is either a bond or
a  path of bonds, joining intervening agents, connecting them 
so that their social relation can be estimated by agent $i$.
We will assume that the representation is a connected graph, what can be accomplished
by defining the distance of two unconnected vertices to be infinite. 
It is reasonable to assume that this lack 
of direct knowledge will imply in a social cost which increases with
the length of the shortest path between the agents. This implements
the idea that relying on heuristics to infer the relationship between them 
(e.g., ``a friend of an enemy is an enemy'', etc.) is more amenable to 
errors as the number of intermediate agents grows.

Call $ l^i_{jk}$ the social distance in the graph defined by the 
adjacency matrix ${\bf S}^i$. The $\lambda-$th power of  ${\bf S}^i$, 
\[\, {\bf M}^i(\lambda)= [{\bf S}^i]^\lambda\]
permits verifying whether there is a path joining two agents, and the social distance
is
\[l^i_{jk}={\tt min} \, \lambda, \mbox{such that  } {\bf M}^i(\lambda)_{jk}>0\]
is the length of the smallest 
path of bonds linking
$j$ and $k$. We take the social cost of agent $i$ of having a representation 
  ${\bf S}^i$ to be just the distance averaged over all pairs of agents
\begin{equation}
\bar{L^i} = \frac{2}{n(n-1)}\sum_{j, k=1}^n l^i_{jk}.
\end{equation} 

The joint
cognitive-social cost of the representation
is defined as a sum of monotonic functions of $N^i_{cog}$ and
  $\bar{L^i}$ and the simplest  form is just
\begin{equation}
\label{eq:cost}
  C_0({\bf S}^i) = N_{cog}^i + \alpha \bar{L^i},             
\end{equation}
For high $\alpha$, optimization is obtained by decreasing  $\bar{L^i}$
independently of $N_{e}^i$.
For low $\alpha$ the number of edges  $N^i_{cog}$ has to be controlled, 
independently of   $\bar{L^i}$. Hence  $\alpha$ is a parameter of the theory that measures the 
relative importance of the social and cognitive components and can
be interpreted as a measure of the cognitive capacity of the agent.

We now attribute a probability $P( {\bf S}^i |I)$ that agent $i$ perceives a network
  $ S^i =\{s^i_{jk}\}$, conditional on any   
available information $I$, using the methods of 
information theory entropic inference. Call ${\cal C}= \E (C_0)$ the 
expected value of $C_0({\bf S}^i)$ under  $P( {\bf S}^i |I)$:
\begin{equation}
{\cal C}= \E (C_0)= \sum_{{\bf S}^i}C_0 P({\bf S}^i|I)
\end{equation}
 Suppose that either
 ${\cal C}$ or equivalently the scale in which fluctuations of $ C_0({\bf S}^i)$ 
above  its minimum are important, are known. Or possibly we just know that
such knowledge would be useful, but we have no access to their specific
values at present.  
The procedure calls for the maximization
of the entropy (see e.g  \cite{ACaticha2008}) subject to the known constraints,
\begin{equation}
P({\bf S}^i|I)={\tt argmax}_P \left\{-\sum_{{\bf S}^i} P({\bf S}^i) \log P({\bf S}^i)
 - \lambda \left[\sum_{{\bf S}^i}  P({\bf S}^i)- 1\right]- \beta \left[\sum_{{\bf S}^i}  P({\bf S}^i)C_0({\bf S}^i) - \cal{C}\right] \right\}
\end{equation}
 The 
result is the standard Boltzmann-Gibbs probability distribution
\begin{equation}
P({\bf S}^i|I) = \frac{1}{Z}e^{-\beta C_0({\bf S}^i) },
\label{boltzmann} 
\end{equation}
where $\beta$ is the Lagrange multiplier conjugated variable to  ${\cal C}$ and 
controls the scale in which the fluctuations are important. 
The information content in 
$\beta$ is equivalent to that in
${\cal C}$. Low $\beta$ values means that ${\bf S}^i$ configurations of high 
joint cost will not be unlikely. For high $\beta$ only configurations near the
ground state will be possible. We can interpret
$\beta$ as an ecological pressure, possibly correlated to a measure of the effort to collect a minimum number of calories in one day. 
The normalization factor  in equation \ref{boltzmann}, the
partition function $Z$, depends on the number of agents, $\alpha$ and $\beta$:
$Z=Z(n,\alpha,\beta)$. 

In order to characterize the state of the system we need appropriate 
order parameters.
In particular we want to probe whether represented
agents are considered symmetrically or if distinctions are made. It is useful
to introduce the degree of  vertex $j$ in a graph ${\bf S}^i$, $d^i_j=\sum_k s^i_{jk}$,
the number of edges 
emerging from the vertex or in the case of the represented social web, the
number of memorized social relations of an agent; as well as the maximum degree and the average 
\[ d^i_{max}={\tt max}_j d^i_j,
\,\,\,\,\,\ d^i_{avg}=\frac{1}{n}\sum_{j=1}^n d^i_j. \]
Natural order parameters are the expectation values   $\E(d_{max})$ 
and $\E(d_{avg})$ with respect to
$P({\bf S}^i|I)$,  the Boltzmann distribution in equation \ref{boltzmann}. 

\section{Methods}
Several techniques can be used to obtain estimates of the order parameters
and here we present results obtained
employing numerical Monte Carlo methods.
We first considered isolated agents and the Monte Carlo simulation
of the Boltzmann distribution (eq \ref{boltzmann}). Then we considered
2-body interactions of the $n$ agents, exchanging information 
about other pairs of individuals, through a
mechanism that can be called gossip. The effect of gossip is to generate
highly correlated perceived social webs.  
The advantage of Monte Carlo methods is that a simple extension of the
type of dynamics permits incorporating very simply interactions like gossip.
 
We now run the simulation for the $n$ agents together. A parameter
$g$ ($0<g<1$) measures the intensity of information exchange through gossip.
Choose an agent $i$ and  pair $(j,k)$, independently of anything else, uniformly
at random. With probability
$1-g$, a MC Metropolis update is performed on the bond of the pair $(j,k)$ .
Let $ \bar{s}^i_{jk} = 1- s^i_{jk}$ be
the complementary value  of  the bond variable  $ s^i_{jk}$.
Also independently and uniformly at random, with probability $g$ another 
agent is chosen, call it $l$. Its corresponding edge   $ s^l_{jk}$ is copied to 
 $ \bar{s}^i_{jk}$. 
Let $\bar C_0 $ be the joint cognitive-social cost with the 
bond $s^i_{jk}$ replaced by  $ \bar{s}^i_{jk}$. 
With probability
$\min\left\{\exp (-\beta (\bar C_0 -C_0)),1\right\}$ let 
the change of $ s^i_{jk}$ by  $ \bar{s}^i_{jk}$ be accepted. Otherwise
 $ s^i_{jk}$ is kept fixed.

 The step performed with probability $1-g$ simulates
the update of the social web representation by independent
observations, learning new relations and forgetting about previously known 
relations. The gossip step, done with probability $g$,
simulates the exchange 
of information where agent $l$ tells and agent $i$ learns or forgets  something
about the relation of agents $j$ and $k$. Gossip can be introduced
by more elaborate schemes but this is sufficient for our modelling purposes.

\begin{figure}
\center
\includegraphics[width=0.95\textwidth]{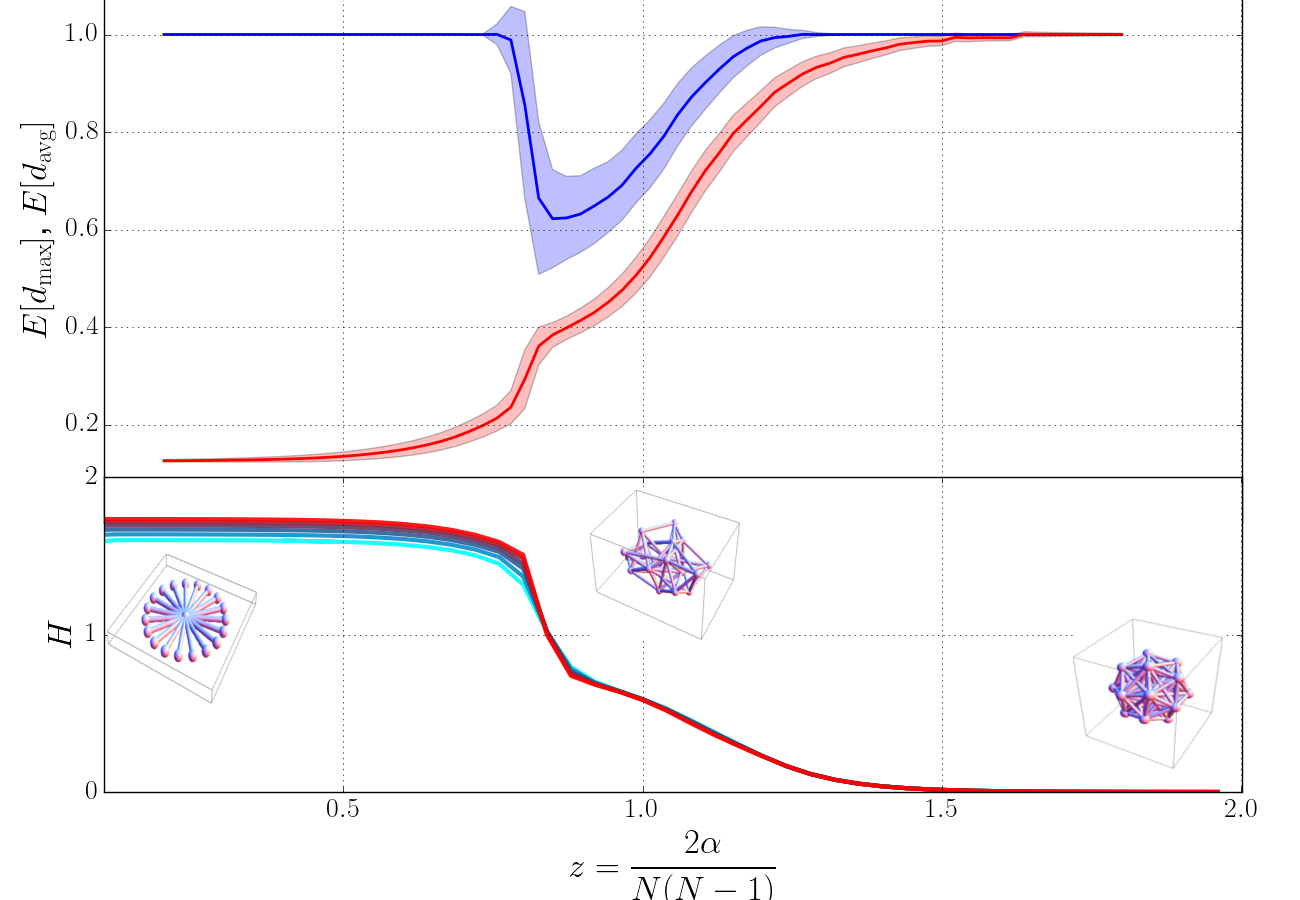}
\caption{Top: Monte Carlo estimates of the maximum degree and the average
degree ($\E(d_{avg})/(n-1)$ and $\E(d_{max})/(n-1)$) of the social web representation as a function of the specific cognitive capacity per dyadic relation $z=2\alpha/n(n-1)$
for $\beta= 10$. The shaded areas are bands at $\pm 1$ std from the Monte Carlo.
 Bottom: $H=2(1-D)$, with $D=\E(d_{avg})/\E(d_{max})$. Roughly three regimes can be seen: 
for very large $z$, $H$ goes to zero (symmetric phase), all agents are equal.
For very small $z$, $H$ is $\approx 2$, the broken symmetry phase, where a
particular agent occupies the central position of the web. An intermediate
$z$ transition region shows intermediate values of $H$. All agents are statistically but not 
strictly equals, some occupy, but only  temporarily, in the stochastic dynamics a more 
central position. 
The different curves are for different values of $n (=10,11,...15)$ , note the almost exact colapse
when plotted as a function of $z$. 
The insets are typical realizations of the inferred web of social interactions by an agent at that $z$ position. Bonds are only drawn if the bond variable is one. 
}
\label{degree}
\end{figure}

\begin{figure}[h]
\center
\includegraphics[scale=0.2, angle=0]{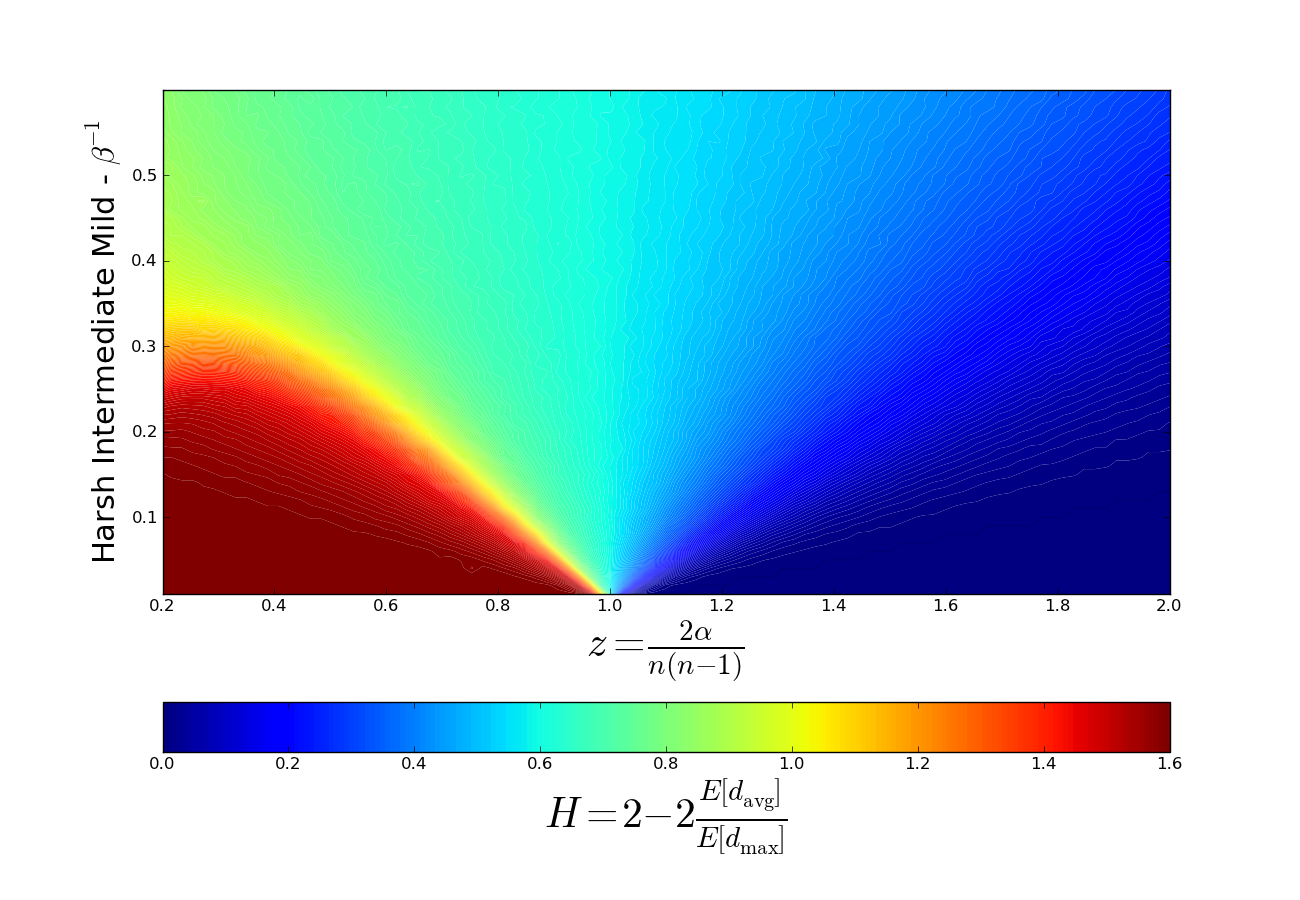}
\includegraphics[scale=0.2, angle=0]{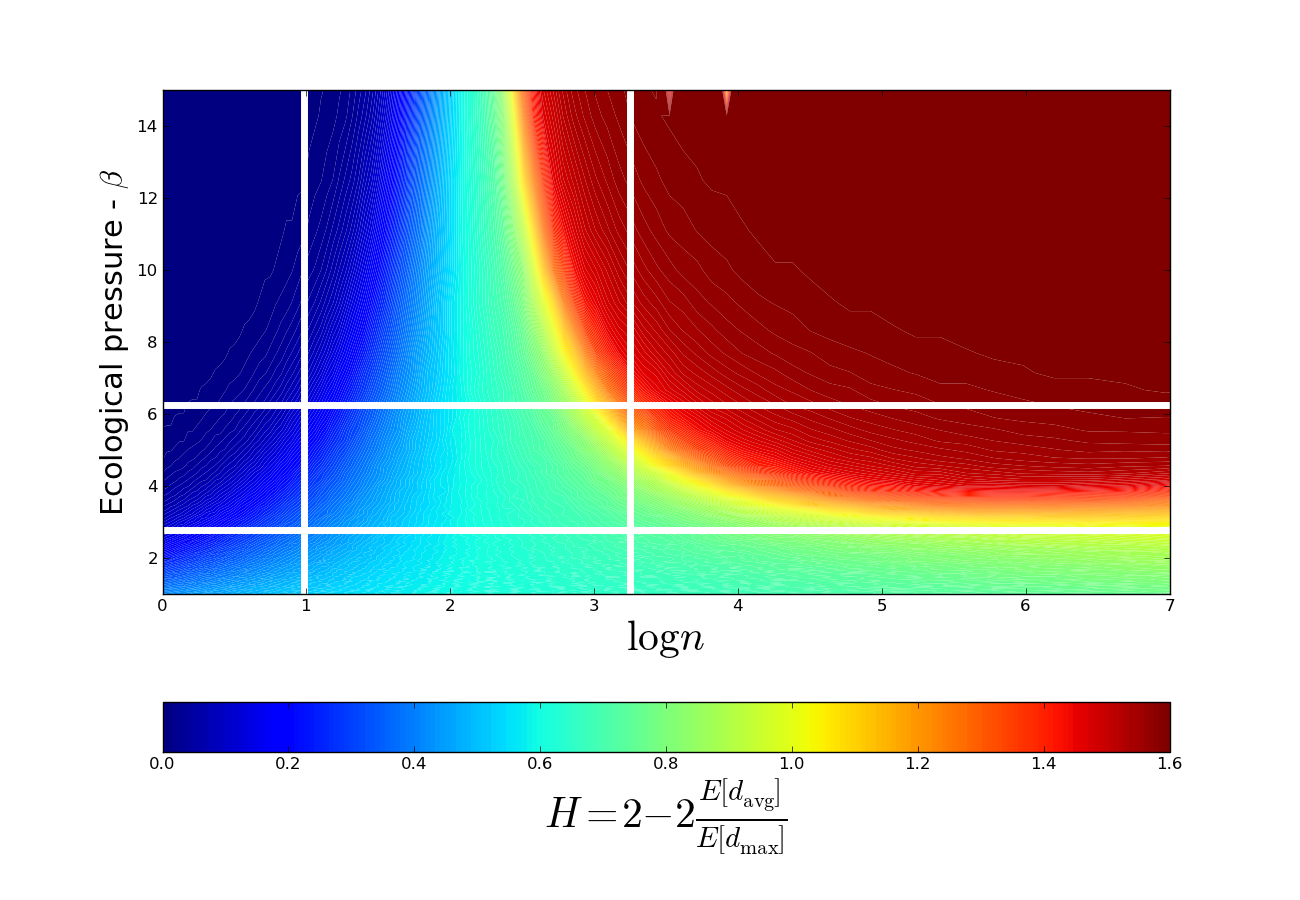}
\caption{Left, Hierarchical-Egalitarian phase transition: 
phase diagram in the plane of specific cognitive capacity per dyadic relation
$z=\frac{2\alpha}{n(n-1)}$  and the inverse ecological pressure 
$\beta^{-1}$. The color code 
represents a measure of the social hierarchy measure or symmetry breaking parameter
$H=2-2\E(d_{avg})/\E(d_{max})$. 
The red region is where the symmetry is broken ($H$ near 2) 
and the maximum degree $\E(d_{max})$ is much larger than the mean 
$\E(d_{avg})$. The blue region is the unbroken symmetry phase, $H\approx 0$. Right,
same but for constant $\alpha$ in the $\log n,\beta$ plane. The white lines divide the phase diagram into 
regions that can be used for the comparison to the Ethnographic Atlas data. The dark red region is where the symmetry is broken and the maximum degree is much larger than the mean. Dark blue is the symmetrical or egalitarian representation region. }
\label{phasediagram}
\end{figure}

After all $i=1,...N$ have been considered, a MC sweep has been completed. 
The $\alpha$ range $4.5 \le \alpha \le 90 $ was divided uniformly into $100$
intervals; the $\beta$ range $0 < \beta \le 20$ was divided into $200$ 
intervals. The values of $n$ varied from 7 to 15. The number of degrees of freedom is $n\frac{n(n-1)}{2} ={\cal O}(n^3)$. We run a MC simulation 
for each fixed $n$ and for each pair of $\alpha$  and $\beta$. 
One to two million MC steps were made for thermalization, and then data 
about the order parameters was collected every 4 MC steps for around four
million MCS.
The results
for the order parameters are shown in figures \ref{degree} and \ref{phasediagram} are discussed below.

The average length of the distances in a given graph ${\bf S}^i$ has to 
be measured in each Metropolis step. It was obtained
using Dijkstra's algorithm to calculate every pair distance.

\begin{figure}[h]
\center
\includegraphics[scale=0.3, angle=0]{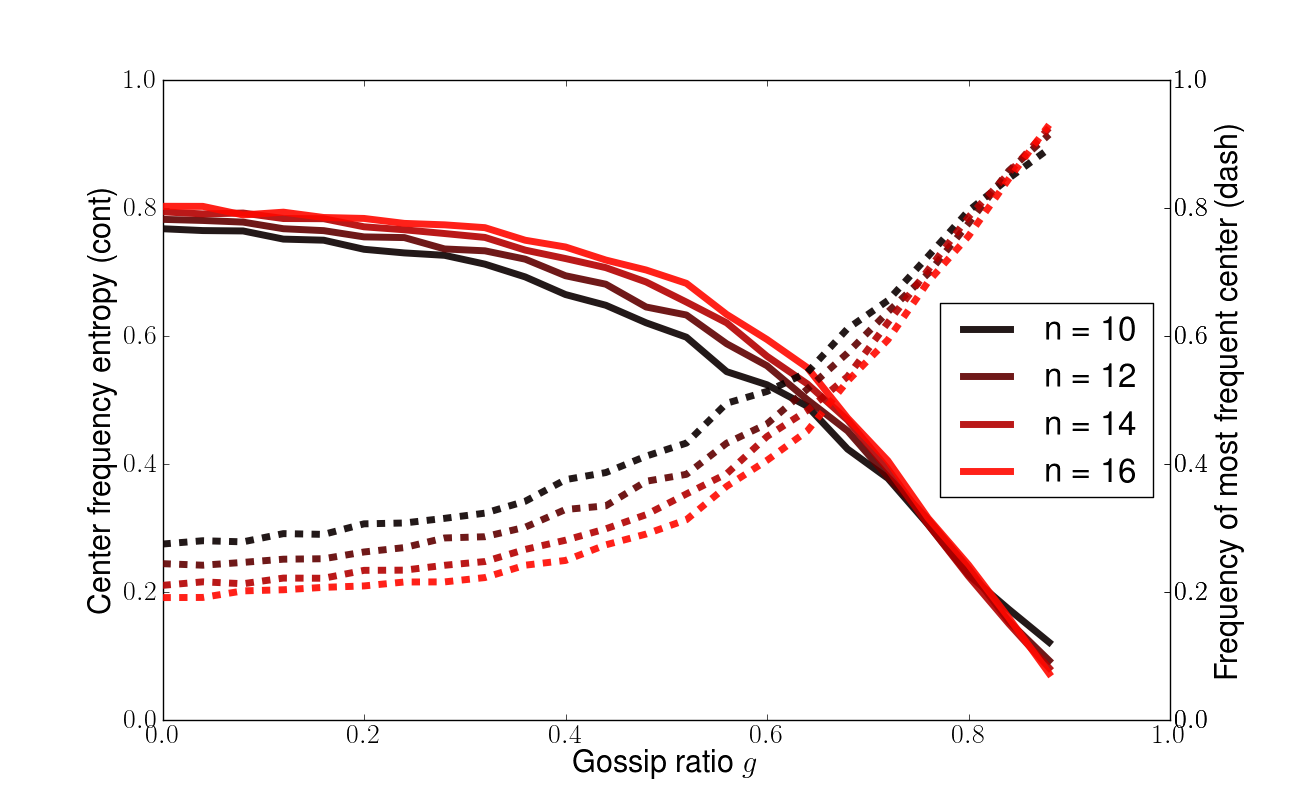}
\caption{The effect of gossip inside the hierarchical phase. 
Continuous curves: The entropy (essentially a 
measure of the width (eq. \ref{entropy2})) of the distribution 
of central agents 
decreases
with the increase of gossip,  meaning that there is a particular agent 
that preferentially occupies the centers of the stars.  
Dashed lines: Another way of seeing the same ordering: 
the frequency of the most frequent 
central agent in their representations.}
\label{gossip}
\end{figure}

{\section{Results}
\subsection{Joint dependence on cognition and band size}
An interesting result is that, given $\beta$, to a good approximation
the properties of the system do not depend separately on $\alpha$ and $n$
but on  the ratio $z=\frac{2\alpha}{n(n-1)}$, which can be thought of as a 
 measure of the effective cognitive capacity per dyadic relation
on a social landscape.

We start by considering the case with no gossip  $g=0$ where agents process
information in a decoupled way. Larger $g$ gives similar results
for the individual
web representations but they are no longer independent and correlations
of the webs appear. 
Figure \ref{degree} shows the results of  Monte Carlo
estimates  of the order parameters,  the expected values of  $d_{max}$
and  $d_{avg}$, respectively$ \E(d_{avg})$ and $\E(d_{max})$.  
These are plotted as a function of the scaling variable
$z=2\alpha/n(n-1)$. 
In the bottom  of figure \ref{degree} 
we show the hierarchical order parameter $H=2-2D$ as a function
 $z$ for $\beta$ fixed, where 
$D=\E(d_{avg})/\E(d_{max})$. 
For $H=0$ the typical graph is  the totally symmetric graph, while for 
$H=2$ the typical graph is the star. Since $n$ is finite, 
$H$ can't be 2. The maximum value is $H_{max}=2(1-2/n+1/n^2)\approx 2$.   
Three  different regimes
can be identified: low, intermediate and high $H$ regions. 
In figure \ref{phasediagram} (left) we show $H$ as a heat map in the $z-\beta^{-1}$ plane. The three phases can be seen again. An intermediate fluid phase has the shape of
a wedge that decreases in width as ecological pressure increases.

\subsection{Gossip and shared perception}
Figure \ref{gossip} shows how frequent is
the most frequent central agent as a function of 
the level of gossip $g$.
Let  $P(c=j|i)$ be the probability that 
for the social web representation of agent $i$
the central element is $j$. The spreading of the probability distribution
can be measured by the ratio of its entropy to the maximum possible value 
\begin{equation}
s_{cf}=\frac{\bar{S}}{\log n}=\frac{-1}{n \log n}\sum_{i,j=1}^n P(c=j|i) \log P(c=j|i)
\label{entropy2}
\end{equation}
 The results indicate that large 
correlation occurs when gossip dynamics dominates independent dynamics, 
starting around $g \approx 0.5$.
\begin{figure}[h]
\center
\includegraphics[scale=0.4, angle=0]{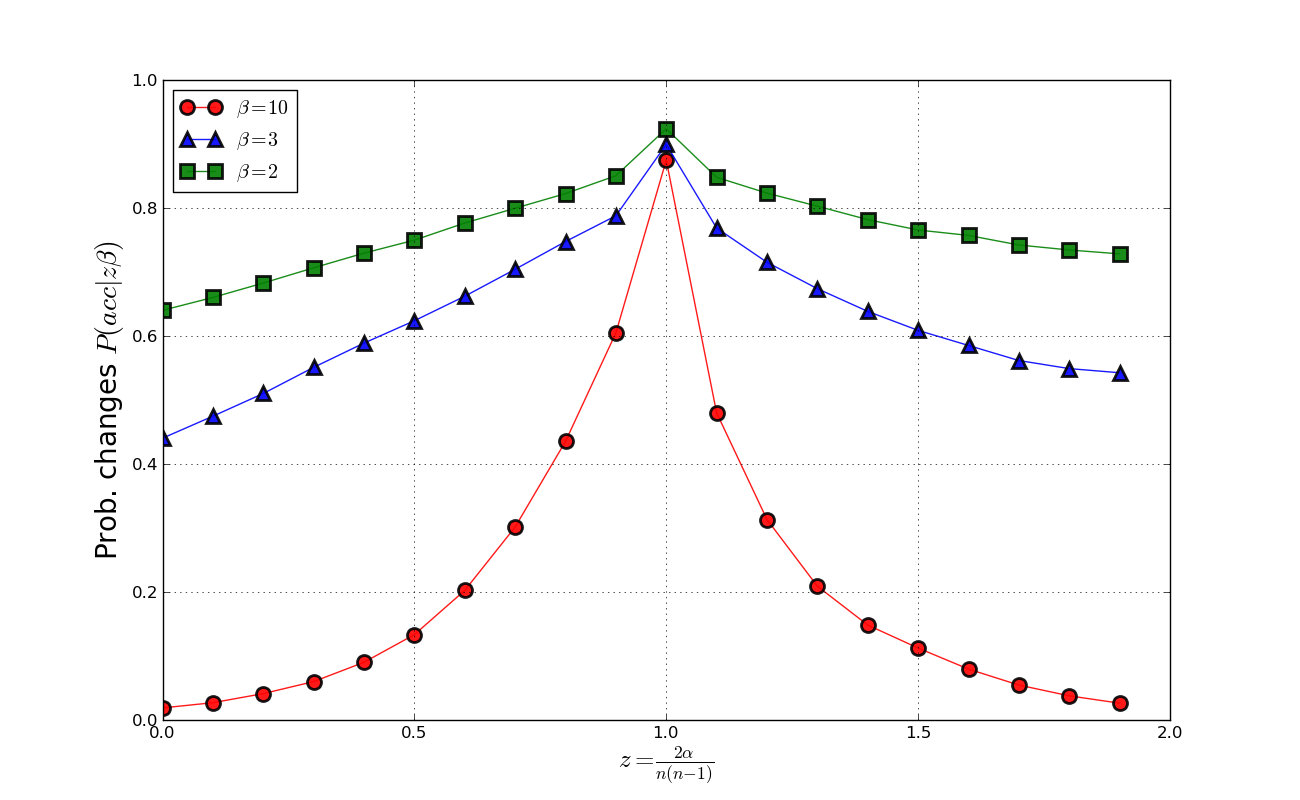}
\caption{The probability of acceptance of changes $P(acc|z \beta)$ from the 
Monte Carlo simulations measures the
tolerance to changes in the social web representations. 
As  a function of
$z$, for different values of $\beta$  ($\beta=2., 3.$ and $10$)
For high pressure, or harsher environments, ($\beta =10$) changes that would permit upstarts
to be different from other agents are not tolerated
for large $z$. This is analogous to counter dominance 
behavior theory
 \cite{Boehm93}. 
Changes of the central agent of the
star topology are unlikely to be accepted for small $z$.
For milder pressures changes are more easily accepted. 
 Tolerance is measured by 
the Monte Carlo acceptance probability 
}
\label{accept}
\end{figure}

Again we stress the hypothesis that the likelihood of an agent in tolerating
inequalities is associated to the perceived inequalities of its social
web representation. The three regimes will have strong influence
in the possibilities of social organization of the group. 
In the region where $H$ is close to zero, the  interpretation is that 
no inequalities can be tolerated. These 
would represent large fluctuations on the cognitive-social cost and the 
combination of cognitive resources and band size given by $z$ is large
enough to permit a representation web given by a full graph. 

The intermediate  wedge region 
could be interpreted as the ``Big Man'' society, where
some inequality is possible, but is not solidified and these temporarily
more central figures can be though of as ``first among equals''  and
their position is liable to changes. Since the wedge decreases
for increasing  pressure, for extreme ecological pressure,
a Big Man organization is
 not possible. Either there is  a stable central figure, 
e.g. a chief, or symmetry among members of the band.

The lower left hand part of figure  \ref{phasediagram} (left)  is where
the symmetry breakdown of the web representation permits the emergence
of tolerance towards inequalities. The exchange of information about
the social webs leads to the choice of an almost unique and stable 
central agent for all agents. 
This would allow the creation of a society where
authority is stable and social egalitarianism is lost. 
In figure \ref{accept} we show the probability of a change in the cognitive 
representation webs as a function of $z$, measured by the Monte Carlo (Metropolis) 
acceptance rate. 
Only in the intermediate Big Man fluid region
a significant rate of changes is acceptable. In both hierarchical and
egalitarian phases, the dynamics turns out to be very 
conservative and change is rare, maintaining status quo for very long times. This prediction of
the theory is in accordance to what is expected from anthropology's Reverse Dominance theory \cite{Boehm93}. 

\subsection{Knauft's U-shape}

\begin{figure}[h]
\center
\includegraphics[scale=0.6, angle=0]{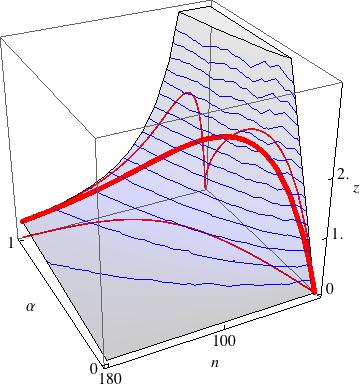}
\caption{Schematic (inverted-) U-shape trajectory for the specific cognitive capacity
$z \propto 2\alpha /n(n-1)$ as a function of time 
(thick curve). The higher the value of $z$, the more symmetrical 
or egalitarian the society will be.
This is just a representation of externally caused changes in 
the cognitive capacity $\alpha(t)$ and the mean size of social groups $n(t)$
as  a function of time. The thin lines are the shadows projected onto the 
respective planes. The surface is $z(\alpha,n) \propto \frac{2\alpha}{n(n-1)}$.
The contours are drawn for constant $z$ values.}
\label{ushape}
\end{figure}

The fact that the phase diagram can be drawn using the combination 
$z=2\alpha/n(n-1)$ immediately suggests a scenario that accommodates
the U-shape dynamics along the egalitarian-hierarchical spectrum.
The schematic drawing in figure \ref{ushape}
shows the curve  $z=z(\alpha(t), n(t))$ in a  parametric representation 
using some rough measure of time as the parameter. We use a simple model 
of the growth of the cognitive capacity $\alpha$ in an evolutionary 
time scale and the fast increase in band sizes $n$ in the transition
to the neolithic. For viewing purposes only we use different
time scales along the trajectory so that the shape is clearly a nice  inverted U, otherwise it would be very skewed, since it takes around 7 million years to
go up from hierarchical to egalitarian and few thousands years to go
down back to hierarchical. It starts with low $z$ around 7 Mya, in the
hierarchical region of  left hand side of the phase diagram of figure \ref{phasediagram}. It 
slowly grows, reaching a peak of $z$ in an egalitarian region due
to increased encephalization. 
Finally it goes back to the region of low $z$ due to increase in 
band size, in the hierarchical phase of figure \ref{phasediagram}. 
Of course the specific details of such  trajectory would
depend on many other conditions, but this furnishes a 
plausible qualitative scenario for the evolution of $z$.

\section{ Ethnographic  data and theoretical  predictions}

\begin{figure}[h]
\center
\includegraphics[scale=0.3, angle=0]{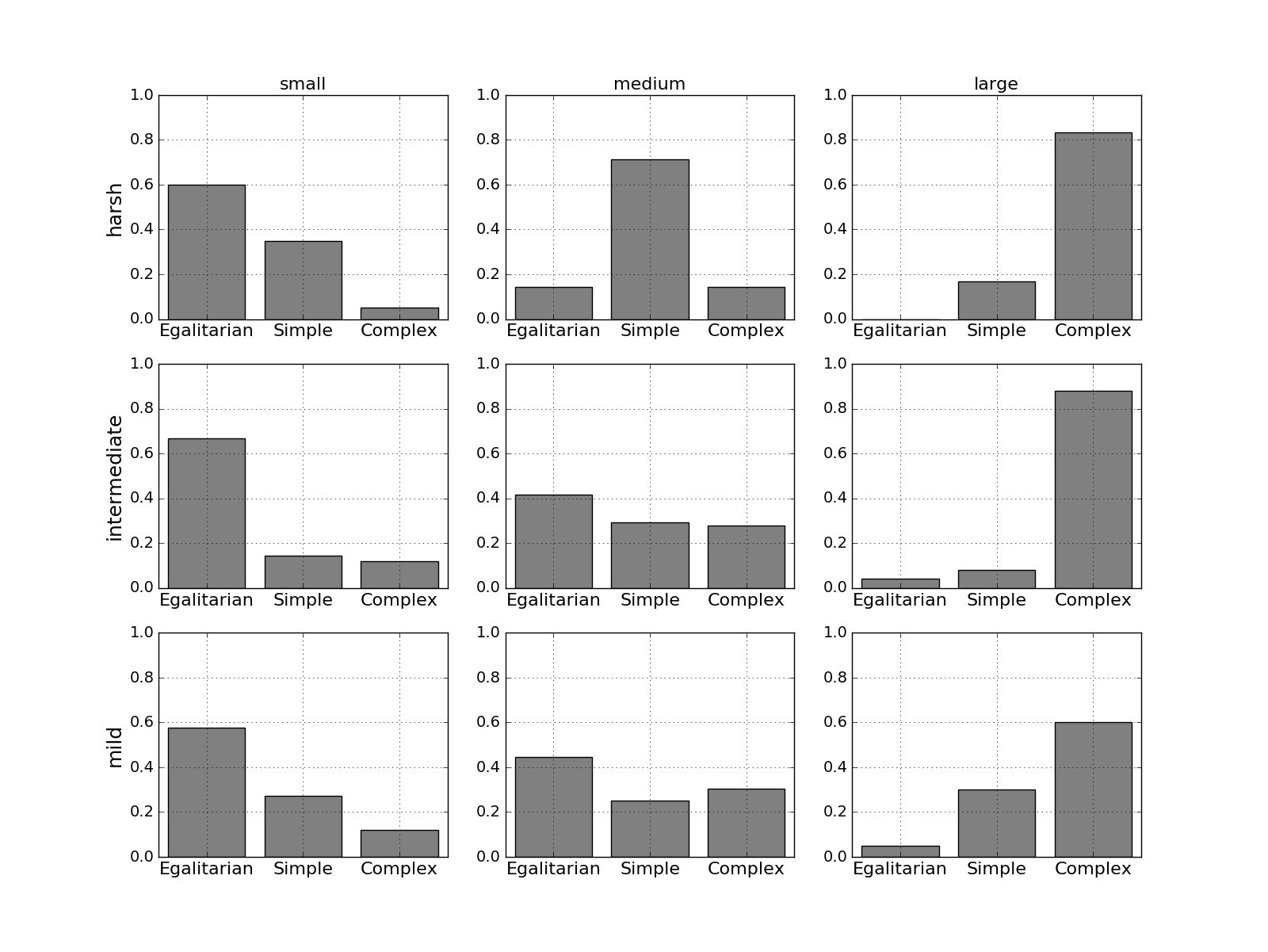}
\caption{}
\label{conditional}
\end{figure}

\begin{figure}
\center
\includegraphics[width=0.95\textwidth]{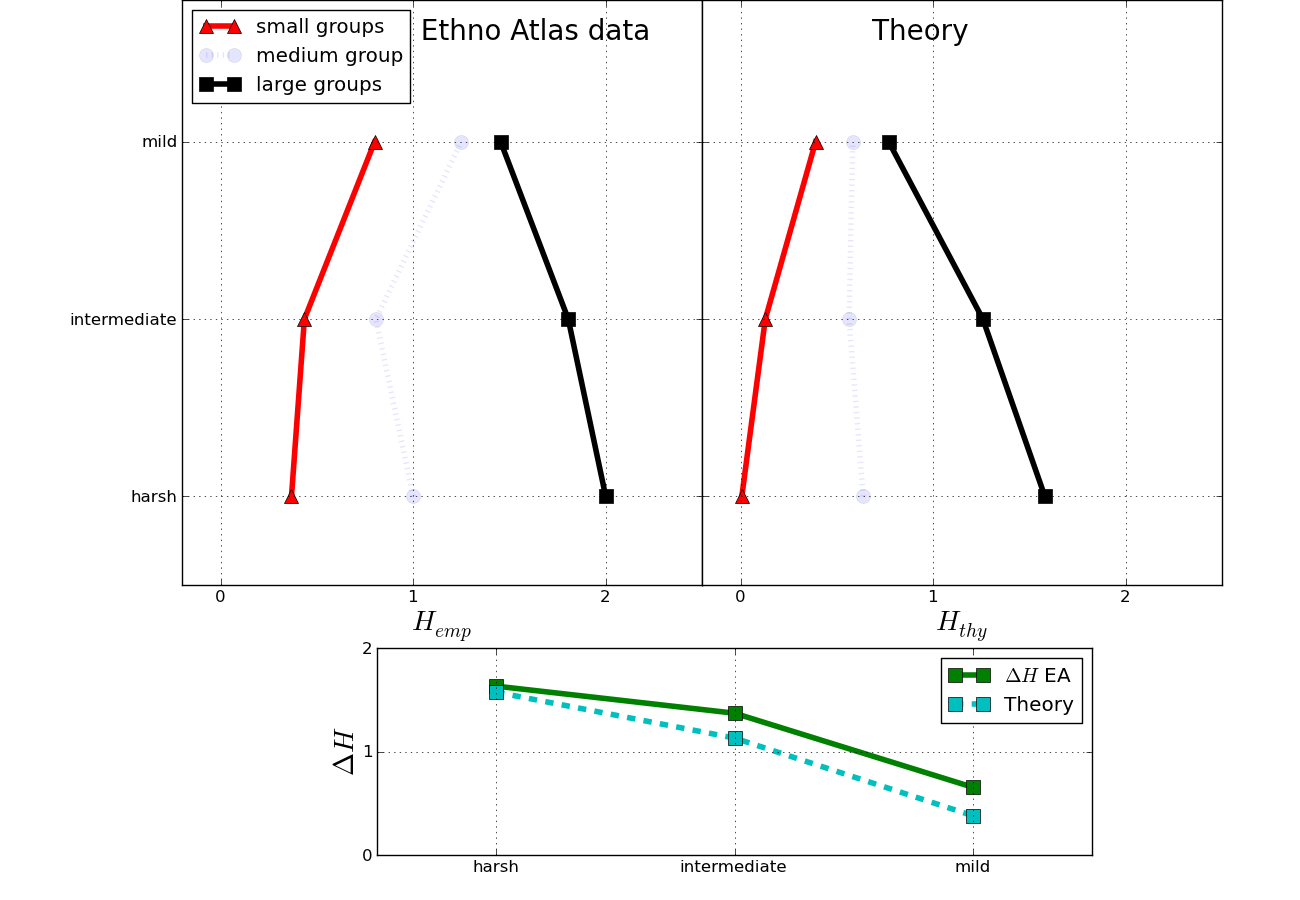}
\caption{The bifurcation signature of the phase transition. 
For  mild climates the expected hierarchies change little with group size.
For harsh climates the expected hierarchy
is larger for large groups and smaller for smaller groups. 
Top Left Ethnographic Atlas data. Top Right: theory.
Bottom: The difference in expected hierarchy $\Delta H$
between large and small groups decreases for milder
climates. Continuous line: EA data, Dashed line: theory. 
Harsh climates are 
Tundra (northern areas), Northern coniferous forest,
High plateau steppe, Desert (including arctic). Mild:
Temperate forest,Temperate grasslands,
Mediterranean,
Oases and certain restricted river valleys.
 }
\label{hmedio}
\end{figure}
Can a signature of this competition 
between cognitive and social navigation constraints 
be seen today for modern humans? 
A clear theoretical prediction about the dependence of social
stratification on ecological pressure $\beta$ and group size ($z=z(n)$,
$\alpha$ fixed) can 
be confronted to data from the ethnographic record.
The  prediction is divided into two 
parts. 
First, for very mild climates intermediate social structures are expected,
but, as climates of
increasing harshness are considered, 
different social organizations will occur.
Second, this difference depends on group size. Cultures organized in
small groups will be  more egalitarian, those in large groups more 
hierarchical.

Using Murdock's Ethnographic Atlas (\cite{ethnoatlas})  we see in
figure \ref{hmedio} that this
prediction is indeed borne out by the data. These are not predictions 
about a specific group becoming more or less hierarchical as climate changes. 
These are predictions about the expectations we should have about
hierarchical organization as different climates and group sizes
are considered. Changes in a particular group would not be so easily observed,
since the shift of perception of social webs will have an influence on motivations. See \cite{Wiessner2002} for a description of a system in the
process of transition. How changing motivations  lead to cultural changes
and influence social organizations is
outside the scope of the present theory.

From the Ethnographic Atlas we extracted
the relevant variables: 
data for social stratification $h$, climate $c$ and group size $s$. Each variable range is divided into three regimes, 
low, intermediate and high ($0,1,2$ respectively, see \cite{SM}). The number of 
cultures in \cite{ethnoatlas} with information on those three 
variables is 248.
From the data we  obtained the
conditional probability $P(h|cs)=P(hcs)/P(cs)$.
The empirical expected hierarchy value
$H_{emp} = \sum_h h P(h|cs)$ for each possible combination of 
climate and size, is shown 
in the left panel of Figure \ref{hmedio}. 
The climate variables describe the type of environment but we need a quantitative description of the climate
instead of just its name. We decided that a reasonable conversion could be done by using the idea
of Net Primary Production \cite{Ecology} which is a measure of the amount of calories per day
that can be extracted from the environment and therefore correlates negatively to the ecological pressure
$\beta$.

To obtain the analogous theoretical predictions, the same
is done by dividing the 
range of theoretical parameters into three intervals as well and
calculating the same quantity from the theoretical results, see figure \ref{phasediagram}  (right).
The theoretical expected value of $H_{thy}=2-2\E(d_{avg})/\E(d_{max})$
 is shown in the right panel of Figure \ref{hmedio} .
  
The qualitative agreement between theory and empirical record supports that our
methodology is capable of suggesting new ways of  looking at the 
available ethnographic records, which can  now come under scrutiny by 
the community of quantitative ethnography dealing with cross-cultural 
studies.  

The fact that inequality rises with group size 
 and that there are ecological factors involved, has been 
previously considered
\cite{Carneiro,Bingham99,Fletcher,Summers2005,Dubreuilbook,Keeley88}, 
but not how the rise is modulated by ecological pressure nor the 
hypothesis that this is due to the competition of cognitive and social 
navigation needs and therefore the influence of climatic pressure 
on hierarchy can be reversed by demographics. This papers are of a theoretical nature in the
spirit of the social sciences. Mathematical  attempts at modeling  are typically absent and at most
show the results of regressions between pairs
of variables extracted from the ethnographic data. 
\section {Discussion}
Our Statistical Mechanics approach, based on entropic inference through maximum entropy methods, 
is a methodological approach to the
mathematical-physics modeling of systems
that incorporates conditioning factors, in this case demographic, 
ecological, social and cognitive.

Our main hypothesis 
is that a social-cognitive cost is relevant to characterize 
probabilistically the perceived social webs. 
The introduction of the 
conjugated parameter $\beta$, with the same informational
content of the average cost is  an unavoidable  theoretical 
consequence.  It controls the size of fluctuations  above the
minimum possible value of the cost, prompting its
interpretation as a pressure. Gossip, a metaphor for 
information exchange, correlates the perceived webs. 
The cognitive capacity and the size of the group combine into 
a variable $z$, the  {\it specific} cognitive capacity,
and the  perceived social state  can be described in a space of just
two dimensions  $(z,\beta)$. 
External to the model, the dynamics of
encephalization and band size, determine the historical evolution 
of $z$ leading to a scenario for non-monotonic hierarchical 
change \cite{Knauft1991}.
Further changes in $z$ could occur, e.g. due to
technological advances which 
translate into more effective information processing and 
better social navigation. Also
an effective reduction of ecological pressure, following
enhanced productivity can occur. Then a  more
egalitarian perception of the social web will follow.
The PMB hypothesis predicts that
motivations and behaviors will change, but the theory does not 
go into the area of predicting how
behaviors change, nor what institutions will emerge in order to permit 
such behaviors, nor the time scales of these changes.  
Our approach to 
the transition from hierarchical to egalitarian and back dispenses the
issue of whether the hierarchical type of behavior lay dormant
(Rodseth  in \cite{Knauft1991})
and remained present throughout the Pleistocene or whether
the resurgence was due to convergent evolution (e.g. \cite{Smail2008}).
It can be turned on or off by the joint effects of cognitive resources,
social demands, ecology and demography.
These transitions resemble the freezing
or evaporation of water by changing pressure or temperature. 
The possibility of being solid ice is not dormant in water when
it is heated up. At least that is not a useful metaphor.

We can speculate that the time spent in the large $z$ egalitarian phase 
promoted conditions for the fixation of altruistic genes
and the emergence of the "do unto others" ideas since
all are equal under the representation web. It is hard
to imagine the fixation of 
altruistic behavior which arises from
 punishment and collaboration  \cite{BoydRicherson, BGB, Schonmann2012}
in other than 
the symmetric phase, but this should be amenable to 
model construction and analytic studies.

This simple model and the
particular function we have used to represent the 
cognitive-social cost are far from complete. We don't claim
specific numerical validation by confrontation with 
empirical data, in any other way than just a qualitative one. More
sophisticated forms of coalitions, other than dyadic pairing, should
lead to increased richness of the phase diagram, without disrupting the 
rough overall picture. We have also avoided considering gender
issues. Rampant sexual inequalities can exist in an egalitarian
organization of males.
Nevertheless, if competition
between cognitive constraints and social navigation needs indeed occur, then 
phase transitions from  egalitarian to hierarchical perception
follows from general arguments.
It has been argued  \cite{Feinman} that ``in the history of the human species,
there is no more significant transition than the emergence and 
institutionalization of inequality.''
We expect that these methods,
which unify the theoretical analysis of the empirical 
facts behind the scenario for the U-shape dynamics  
and the conditions that influence the transformation of perception of
social organization, 
will stimulate the use of information theory methods in 
the analysis of empirical research in cross-cultural studies. 

{\bf Acknowledgments:} We thank Bruce Knauft for comments on a previous 
version of this paper and Helena L. Caticha for preparing Figure 1. This work was supported by Fapesp (2008/10830-2) and CNAIPS-USP.

\bibliography{authority}

{\appendix}
\section{Appendix}
\subsection{Theory: Conditional Probabilities and order parameters}
The phase diagram in the 
$\beta-z$ is shown right figure in panel \ref{phasediagram}.
We divided the ranges of $\beta$ and $z$ into three regions each:
harsh, intermediate and mild climates and small medium and large
groups respectively. The phase diagram is thus divided into 9 regions.
The regions are chosen essentially so that all points in the 
$\beta-z$ space in the harsh-large region are of the same color
(blue).
The same is done for the region of harsh-small (all red)
and for mild-small and mild-large. The white lines show a reasonable 
choice of what is meant by large, intermediate and small both for $\beta$
and $n$.
A reasonable choice for 
the values separating the three regions are $\beta^{-1}_{HI}=0.15$ and
$\beta^{-1}_{IM}=0.35$. For $\beta^{-1} > \beta^{-1}_{IM}$, climate
$c$ is mild.
For $\beta^{-1}_{HI}<\beta^{-1} < \beta^{-1}_{IM}$, climate  $c$ is intermediate,
and  $\beta^{-1}<\beta^{-1}_{HI} $, $c$ is harsh.
 
For $z=\frac{2\alpha}{n(n-1)}$ the borders are set at 
$z_2=0.6$ and $z_1= 1.4$
For $2> z > z_1$, $s=1$ small.
For $z_2>z> z_1 $, $s=2$ intermediate.
For $0<z<z_2$, $s=3$ large.
Then we consider the order parameter $D(s,c)$
\begin{equation}
\bar{D}(s,c)=\frac{\int_{\beta \in c}\int_{z \in s}D dz d\beta}
{\int_{\beta \in c}\int_{z \in s} dz d\beta },
\end{equation}
where 
$
D=\frac{\E(d_{avg})}{\E(d_{max})}.
$ and the theoretical hierarchical order parameter that can be compared to the data is $H_{thy}= 2-2\bar{D}(s,c)$.

\subsection{Data: Source}
Data was obtained from \cite{ethnoatlas}, the 
 Ethnographic Atlas (EA) ``a database on 1167 societies coded by George P. Murdock and published in 29 successive installments in the journal ETHNOLOGY, 1962-1980'', available for download
from the site of Douglas R. White

{\tt 
http://eclectic.ss.uci.edu/~drwhite/worldcul/world.htm}

We used the file {\tt EthnographicAtlasWCRevisedByWorldCultures.sav}

The relevant variables for our study are $s$ , $h$ and $c$, which stand for 
size category, hierarchy category and climate category. All
variables can take integer values  $1$, $2$ or $3$. They are obtained by grouping the EA variables into three groups:

\begin{table}[h]\small 
\begin{tabular}{|l|c|c|c|} 
\hline
$\downarrow$Category, Value $\rightarrow$
 & 1 & 2 & 3 \\
\hline
s: group size (v31) &  small & medium  &  large\\
h: social stratification (v66) & Egalitarian  & Simple structure  & Complex \\
c: Climate (v95)        &  Harsh & Intermediate &  Mild \\
\hline
\end{tabular}
\caption{EA variables and categories}
\end{table}

\begin{table}[h]\small 
\begin{tabular}{|l|l|c|c|c|} 
\hline
&&Number of cultures in&Number of cultures in&Number of cultures in\\
$\downarrow$Stratification&Climate & Small groups & Medium groups& Large groups \\
\hline
 1& 1& 12& 1 &0\\
 1& 2& 43& 40& 2\\
 1& 3& 4 &3 &0\\
 2& 1& 7 &5 &0\\
 2& 2& 11& 25& 3\\
 2& 3& 4 &3 &6\\
 3& 1& 0 &1 &3\\
 3& 2& 8 &23& 31\\
 3& 3& 2 &6 &5\\
\hline
\end{tabular}
\caption{Number of Cultures in the different categories in the EA.}
\end{table}

\subsection{Data: Conditional Probabilities and order parameters}
 This values
are obtained by grouping the relevant variables of the EA
according to tables 1-3SM below, into three categories. The 
results are presented in table 4SM below.
 We extract the numbers of cultures
$N(s,h,c)$ 
with a given set of values $(s,h,c)$ and the marginal numbers
$N(s,c)$ of cultures with a given pair of values of $(s,c)$
independently of $h$. These are related by 
$N(s,c)=\sum_{h=1,2,3}N(s,h,c)$. 
The conditional probabilities are
\begin{equation}
P(h|sc)=\frac{N(s,h,c)}{N(s,c)},
\end{equation}
of a culture having a given class stratification , given 
its climate and group size. 

Then we calculate the average hierarchy of the cultures with the
same values of $n$ and $c$, that is, that belong to the same
size and climate categories. We calculate the empirical average hierarchies
conditional on size and climate 

\begin{equation}
\bar{H}=\E(h-1|sc)= \sum_{h=1,2,3} (h-1) P(h|sc),
\end{equation}
which satisfies $0 \le \bar{H} \le 2$.

Fluctuations around the average 
\begin{equation}
\sigma_{EA}^2=\E((h-\bar{h})^2|nc)= \sum_{h=1,2,3}(h-\bar{h})^2P(h|nc),
\end{equation}
can be calculated to define error bars.



\newpage

\subsection{Numerical results}

\begin{table}[h]\small 
\begin{tabular}{|l|l|c|c|} 
\hline
$\downarrow$Climate, Group size $\rightarrow$ & Small & Medium & Large groups \\
\hline
Harsh &   .37    &     1. & 2.  \\
Intermediate & .44  &  .81 & 1.81  \\
Mild         &  .80  & 1.25 &  1.45 \\
\hline
\end{tabular}
\caption{The results for the empirical stratification $\bar{H}$}
\end{table}
\begin{table}[h]\small 
\begin{tabular}{|l|l|c|c|} 
\hline
$\downarrow$Climate, Group size $\rightarrow$ & Small & Medium & Large groups \\
\hline
Harsh &      .01 &     .64 &      1.58\\
Intermediate &  .13 & .56 & 1.25\\
Mild         &  .39 & .58 & .77\\
\hline
\end{tabular}
\caption{The results for the theoretical prediction $H_T$}
\end{table}

\newpage

\subsection{Ethnographic data}

\begin{table}[h]\small 
\begin{tabular}{|r|l|c|c|} 
\hline
N & Code & Description v31. & Size category \\
\hline
681 & 0& Missing data (code .)& 0\\ 
118 &1& Fewer than 50 & 1\\
107 &2& 50-99 & 1\\
104 &3& 100-199& 2\\
83 &4& 200-399 &2\\
60 &5& 400-1000& 2\\
16 &6& 1,000 without any town of more than 5,000 & 3\\
36 &7& Towns of 5,000-50,000 (one or more) & 3\\
62 &8& Cities of more than 50,000 (one or more)& 3\\
\hline
\end{tabular}
\caption{Variable v31 of the Ethnographic Atlas: Mean Size of Local Communities.}
\end{table}

\begin{table}[h]\small 
\begin{tabular}{|r|l|c|c|} 
\hline
N & Code & Description v66. & Hierarchy category \\
\hline
182 &0& Missing data (code .) & 0 \\
533 &1& Absence among freemen (O.)& 1\\ 
206 &2& Wealth distinctions (W.) & 2\\
39  &3& Elite (based on control of land or other resources (E.) &2\\
228 &4& Dual (hereditary aristocracy) (D.) &3\\
79  &5& Complex (social classes) (C.) &3\\
\hline
\end{tabular}
\caption{Variable v66 of the Ethnographic Atlas: Class Stratification.}
\end{table}

\newpage
\begin{table}[h]\small 
\begin{tabular}{|r|l|c|c|} 
\hline
N & Code & Description v95. & Climate category \\
\hline
869& 0& Not coded & 0\\
3 &51 &Desert (including arctic) &1\\
11 &23& Tundra (northern areas) & 1\\
21 &36& Northern coniferous forest &1\\
8 &44 &High plateau steppe &1\\
5 &65 & Oases and certain restricted river valleys &1\\
37& 52& Desert grasses and shrubs&2 \\
16& 56& Temperate woodland &2\\
24& 74& Sub-tropical bush &2\\
27 &78& Sub-tropical rain forest &2\\
64 &84& Tropical grassland &2\\
14 &87& Monsoon forest &2\\
113& 88& Tropical rain forest&2\\
25& 54& Temperate grasslands &3\\
19 &46& Temperate forest (mostly mountainous) &3\\
11& 55& Mediterranean (dry, deciduous, and evergreen forests) &3\\
\hline
\end{tabular}
\caption{v95 Climate: Primary Environment. Group 1 is formed by NPP up to $350 gC/m^2/year$
Group 2: between $350 gC/m^2/year$ and $600 gC/m^2/year$. Group 3: large than $600 gC/m^2/year$.}
\end{table}
\newpage
\subsection{Cultures: Size, Class Stratification , Climate}
\vspace{5cm}
{\bf Table 8} List of all cultures with available information in 
all three categories 
\begin{table}[h]\footnotesize
\begin{tabular}{|r|l|c|c|c|} 
\hline
 \,&Culture               &    Size(v31)  &     Stratification(v66) &    
Climate(v95)  \\
\hline
  1 & !KUNG                 &       1 &       1 &       2  \\
  2 & ILA                  &       2 &       2 &       2  \\
  3 & NYORO                &       2 &       3 &       2  \\
  4 & AMBA                 &       2 &       1 &       2  \\
  5 & KPE                  &       1 &       2 &       2  \\
  6 & FON                  &       3 &       3 &       2  \\
  7 & KISSI                &       2 &       1 &       2  \\
  8 & BAMBARA              &       3 &       3 &       2  \\
  9 & YATENGA              &       3 &       3 &       2  \\
 10 & KATAB                &       2 &       1 &       2  \\
 11 & KONSO                &       3 &       2 &       3  \\
 12 & SOMALI               &       1 &       2 &       2  \\
 13 & WOLOF                &       3 &       3 &       2  \\
 14 & TEDA                 &       1 &       3 &       3  \\
 15 & BARABRA              &       1 &       2 &       1  \\
 16 & GHEG                 &       2 &       1 &       1  \\
 17 & NEWENGLAN            &       3 &       3 &       2  \\
 18 & DUTCH                &       3 &       3 &       2  \\
 19 & SERBS                &       3 &       3 &       2  \\
 20 & SYRIANS              &       3 &       2 &       3  \\
 21 & SINDHI               &       3 &       2 &       2  \\
 22 & KAZAK                &       1 &       3 &       3  \\
 23 & GILYAK               &       1 &       1 &       1  \\
 24 & YAKUT                &       1 &       2 &       1  \\
 25 & KOREANS              &       3 &       3 &       2  \\
 26 & LOLO                 &       2 &       3 &       3  \\
 27 & ABOR                 &       2 &       2 &       2  \\
 28 & CHENCHU              &       1 &       1 &       2  \\
 29 & TAMIL                &       3 &       3 &       2  \\
 30 & ANDAMANES            &       1 &       1 &       2  \\
 31 & MERINA               &       3 &       3 &       2  \\
 32 & GARO                 &       2 &       2 &       2  \\
 33 & LAMET                &       1 &       2 &       2  \\
 34 & MNONGGAR             &       2 &       2 &       2  \\
 35 & ATAYAL               &       2 &       1 &       2  \\
 36 & SAGADA               &       3 &       2 &       2  \\
 37 & JAVANESE             &       3 &       3 &       2  \\
 38 & MACASSARE            &       2 &       3 &       2  \\
 39 & ARANDA               &       1 &       1 &       2  \\
 40 & KAPAUKU              &       1 &       2 &       2  \\
\end{tabular}
\end{table}
\begin{table}[h]\footnotesize
\begin{tabular}{|r|l|c|c|c|} 
 \,&Culture               &    Size(v31)  &     Stratification(v66) &    
Climate(v95)  \\
\hline
 41 & WANTOAT              &       1 &       1 &       2  \\
 42 & TRUKESE              &       2 &       1 &       2  \\
 43 & TROBRIAND            &       2 &       3 &       2  \\
 44 & SAMOANS              &       1 &       3 &       2  \\
 45 & TIKOPIA              &       2 &       3 &       2  \\ 
46 & NABESNA              &       1 &       1 &       1  \\
 47 & TAREUMIUT            &       2 &       2 &       1  \\
 48 & TWANA                &       1 &       2 &       1  \\
 49 & NOMLAKI              &       2 &       2 &       3  \\
 50 & TENINO               &       2 &       2 &       1  \\
 51 & OJIBWA               &       1 &       1 &       1  \\
 52 & HURON                &       2 &       2 &       1  \\
 53 & HANO                 &       2 &       1 &       2  \\
 54 & CUNA                 &       1 &       2 &       2  \\
 55 & WARRAU               &       1 &       1 &       2  \\
 56 & MUNDURUCU            &       1 &       1 &       2  \\
 57 & SIRIONO              &       1 &       1 &       2  \\
 58 & TUCUNA               &       2 &       1 &       2  \\
 59 & INCA                 &       3 &       3 &       1  \\
 60 & YAHGAN               &       1 &       1 &       1  \\
 61 & MATACO               &       1 &       1 &       2  \\
 62 & TRUMAI               &       1 &       1 &       2  \\
 63 & DOROBO               &       1 &       1 &       2  \\
 64 & NAMA                 &       2 &       2 &       2  \\
 65 & LOZI                 &       1 &       3 &       2  \\
 66 & BEMBA                &       2 &       3 &       2  \\
 67 & KUBA                 &       2 &       3 &       2  \\
 68 & CHAGGA               &       2 &       3 &       2  \\
 69 & KIKUYU               &       2 &       2 &       2  \\
 70 & FANG                 &       1 &       2 &       2  \\
 71 & ASHANTI              &       3 &       3 &       2  \\
 72 & DOGON                &       2 &       2 &       2  \\
 73 & TALLENSI             &       2 &       2 &       2  \\
 74 & TIV                  &       2 &       1 &       2  \\
 75 & AZANDE               &       2 &       3 &       2  \\
 76 & MASAI                &       1 &       1 &       2  \\
 77 & TIGRINYA             &       3 &       3 &       2  \\
 78 & SONGHAI              &       3 &       3 &       2  \\
 79 & SIWANS               &       3 &       2 &       3  \\
 80 & EGYPTIANS            &       3 &       3 &       3  \\
\end{tabular}
\end{table}
\newpage
\begin{table}[h]\footnotesize
\begin{tabular}{|r|l|c|c|c|} 
 \,&Culture               &    Size(v31)  &     Stratification(v66) &    
Climate(v95)  \\
\hline
 81 & RIFFIANS             &       3 &       2 &       3  \\
 82 & ROMANS               &       3 &       3 &       3  \\
 83 & IRISH                &       3 &       3 &       2  \\
 84 & LAPPS                &       1 &       2 &       1  \\
 85 & HUTSUL               &       3 &       2 &       3  \\
 86 & PATHAN               &       2 &       3 &       2  \\
 87 & KHALKA               &       1 &       3 &       2  \\
 88 & CHUKCHEE             &       1 &       2 &       1  \\
 89 & YURAK                &       1 &       2 &       1  \\
 90 & MIAO                 &       2 &       1 &       2  \\
 91 & BURUSHO              &       2 &       3 &       1  \\
 92 & LEPCHA               &       2 &       2 &       3  \\
 93 & BENGALI              &       3 &       3 &       2  \\
 94 & MARIAGOND            &       1 &       2 &       2  \\
 95 & TODA                 &       1 &       1 &       2  \\
 96 & TANALA               &       2 &       3 &       2  \\
 97 & VEDDA                &       1 &       1 &       2  \\
 98 & BURMESE              &       3 &       3 &       2  \\
 99 & SEMANG               &       1 &       1 &       2  \\
100 & ANNAMESE             &       3 &       3 &       2  \\
101 & IFUGAO               &       2 &       2 &       2  \\
102 & SUBANUN              &       1 &       1 &       2  \\
103 & BALINESE             &       2 &       3 &       2  \\
104 & ALORESE              &       2 &       2 &       2  \\
105 & MURNGIN              &       1 &       1 &       2  \\
106 & TIWI                 &       2 &       1 &       2  \\
107 & WOGEO                &       1 &       1 &       2  \\
108 & MAJURO               &       2 &       3 &       2  \\
109 & IFALUK               &       1 &       1 &       2  \\
110 & KURTATCHI            &       2 &       3 &       2  \\
111 & LESU                 &       2 &       1 &       2  \\
112 & BUNLAP               &       1 &       2 &       2  \\
113 & LAU                  &       1 &       2 &       2  \\
114 & PUKAPUKAN            &       2 &       1 &       2  \\
115 & MAORI                &       2 &       3 &       3  \\
116 & MARQUESAN            &       1 &       3 &       2  \\
117 & COPPERESK            &       1 &       1 &       1  \\
118 & KASKA                &       1 &       1 &       1  \\
119 & YUROK                &       1 &       2 &       3  \\
120 & TUBATULAB            &       1 &       1 &       2  \\
\end{tabular}
\end{table}
\begin{table}[h]\footnotesize
\begin{tabular}{|r|l|c|c|c|} 
 \,&Culture               &    Size(v31)  &     Stratification(v66) &    
Climate(v95)  \\
\hline
121 & HAVASUPAI            &       2 &       1 &       2  \\
122 & SANPOIL              &       1 &       1 &       3  \\
123 & OMAHA                &       1 &       1 &       3  \\
124 & CREEK                &       2 &       1 &       3  \\
125 & NAVAHO               &       2 &       1 &       2  \\
126 & ZUNI                 &       3 &       1 &       2  \\
127 & AZTEC                &       3 &       3 &       3  \\
128 & BARAMACAR            &       1 &       1 &       2  \\
129 & TAPIRAPE             &       2 &       1 &       2  \\
130 & JIVARO               &       1 &       1 &       1  \\
131 & YAGUA                &       1 &       1 &       2  \\
132 & AYMARA               &       2 &       2 &       1  \\
133 & CAYAPA               &       2 &       1 &       2  \\
134 & MAPUCHE              &       1 &       2 &       3  \\
135 & BACAIRI              &       1 &       1 &       2  \\
136 & NAMBICUAR            &       1 &       1 &       2  \\
137 & AWEIKOMA             &       1 &       1 &       2  \\
138 & RAMCOCAME            &       2 &       1 &       2  \\
139 & MBUTI                &       2 &       1 &       2  \\
140 & MBUNDU               &       2 &       3 &       2  \\
141 & VENDA                &       2 &       3 &       3  \\
142 & NYAKYUSA             &       2 &       1 &       2  \\
143 & MENDE                &       2 &       3 &       2  \\
144 & YORUBA               &       3 &       3 &       2  \\
145 & BIRIFOR              &       2 &       1 &       2  \\
146 & MAMBILA              &       2 &       1 &       3  \\
147 & MARGI                &       2 &       1 &       2  \\
148 & MAMVU                &       1 &       1 &       2  \\
149 & SHILLUK              &       2 &       3 &       2  \\
150 & LANGO                &       2 &       2 &       2  \\
151 & IRAQW                &       2 &       2 &       2  \\
152 & MZAB                 &       3 &       3 &       3  \\
153 & KABYLE               &       2 &       1 &       2  \\
154 & TRISTAN              &       2 &       1 &       3  \\
155 & WALLOONS             &       3 &       3 &       2  \\
156 & CZECHS               &       3 &       3 &       2  \\
157 & HEBREWS              &       3 &       3 &       3  \\
158 & HAZARA               &       2 &       2 &       2  \\
159 & KORYAK               &       2 &       2 &       1  \\
160 & YUKAGHIR             &       1 &       1 &       1  \\
\end{tabular}
\end{table}
\begin{table}[h]\footnotesize
\begin{tabular}{|r|l|c|c|c|} 
\,&Culture               &    Size(v31)  &     Stratification(v66) &    Climate(v95)  \\
\hline
161 & JAPANESE             &       3 &       3 &       2  \\
162 & MINCHINES            &       3 &       3 &       2  \\
163 & TIBETANS             &       3 &       3 &       1  \\
164 & COORG                &       2 &       3 &       2  \\
165 & KERALA               &       3 &       3 &       2  \\
166 & NICOBARES            &       1 &       1 &       2  \\
167 & SINHALESE            &       3 &       3 &       2  \\
168 & KACHIN               &       2 &       3 &       3  \\
169 & PURUM                &       1 &       2 &       2  \\
170 & CAMBODIAN            &       3 &       3 &       2  \\
171 & HANUNOO              &       2 &       1 &       2  \\
172 & DUSUN                &       2 &       2 &       2  \\
173 & DIERI                &       1 &       1 &       2  \\
174 & KARIERA              &       1 &       1 &       2  \\
175 & KERAKI               &       1 &       1 &       2  \\
176 & PONAPEANS            &       1 &       3 &       2  \\
177 & YAPESE               &       1 &       3 &       2  \\
178 & ULAWANS              &       2 &       1 &       2  \\
179 & NASKAPI              &       1 &       1 &       1  \\
180 & EYAK                 &       1 &       2 &       1  \\
181 & ATSUGEWI             &       1 &       2 &       3  \\
182 & MIAMI                &       2 &       1 &       2  \\
183 & CHEROKEE             &       2 &       1 &       2  \\
184 & DELAWARE             &       2 &       1 &       2  \\
185 & MARICOPA             &       1 &       1 &       2  \\
186 & TAOS                 &       2 &       1 &       2  \\
187 & HUICHOL              &       2 &       2 &       2  \\
188 & CHOCO                &       1 &       1 &       2  \\
189 & CARINYA              &       2 &       1 &       2  \\
190 & GUAHIBO              &       1 &       1 &       2  \\
191 & CUBEO                &       1 &       1 &       2  \\
192 & TUNEBO               &       2 &       1 &       2  \\
193 & ONA                  &       1 &       1 &       1  \\
194 & CHOROTI              &       1 &       1 &       2  \\
195 & CAMAYURA             &       2 &       1 &       2  \\
196 & BOTOCUDO             &       1 &       1 &       2  \\
197 & SOTHO                &       2 &       3 &       3  \\
198 & YAO                  &       2 &       1 &       2  \\
199 & YOMBE                &       2 &       3 &       2  \\
200 & GANDA                &       3 &       3 &       2  \\
\end{tabular}
\end{table}
\begin{table}[h]\footnotesize
\begin{tabular}{|r|l|c|c|c|} 
\,&Culture               &    Size(v31)  &     Stratification(v66) &    Climate(v95)  \\
\hline
201 & BETE                 &       2 &       1 &       2  \\
202 & NUPE                 &       2 &       3 &       2  \\
203 & CONIAGUI             &       1 &       1 &       2  \\
204 & BAYA                 &       2 &       1 &       2  \\
205 & LUO                  &       2 &       2 &       2  \\
206 & CHEREMIS             &       2 &       2 &       2  \\
207 & NURI                 &       2 &       2 &       2  \\
208 & AINU                 &       1 &       1 &       2  \\
209 & OKINAWANS            &       3 &       3 &       2  \\
210 & DARD                 &       2 &       3 &       2  \\
211 & BHIL                 &       1 &       3 &       2  \\
212 & AKHA                 &       2 &       1 &       2  \\
213 & PAIWAN               &       2 &       3 &       2  \\
214 & WIKMUNKAN            &       1 &       1 &       2  \\
215 & ENGA                 &       2 &       1 &       2  \\
216 & LAKALAI              &       2 &       2 &       2  \\
217 & ATTAWAPIS            &       1 &       1 &       1  \\
218 & DIEGUENO             &       2 &       1 &       2  \\
219 & WASHO                &       1 &       1 &       3  \\
220 & PAWNEE               &       2 &       3 &       3  \\
221 & COCHITI              &       2 &       1 &       2  \\
222 & YUCATECMA            &       3 &       3 &       2  \\
223 & WAICA                &       1 &       1 &       2  \\
224 & TEHUELCHE            &       1 &       1 &       2  \\
225 & NGONI                &       2 &       3 &       2  \\
226 & WUTE                 &       2 &       3 &       2  \\
227 & BRAZILIAN            &       3 &       3 &       2  \\
228 & BULGARIAN            &       3 &       3 &       2  \\
229 & BASSERI              &       2 &       2 &       2  \\
230 & KET                  &       1 &       1 &       1  \\
231 & MINCHIA              &       2 &       2 &       3  \\
232 & KHASI                &       1 &       3 &       2  \\
233 & SIAMESE              &       3 &       2 &       2  \\
234 & PURARI               &       2 &       2 &       2  \\
235 & ONOTOA               &       2 &       2 &       2  \\
236 & MANUS                &       2 &       2 &       2  \\
237 & YUKI                 &       1 &       2 &       3  \\
238 & NATCHEZ              &       2 &       2 &       2  \\
239 & JEMEZ                &       2 &       1 &       2  \\
240 & BLACKCARI            &       3 &       1 &       2  \\
241 & MAM                  &       3 &       2 &       3  \\
242 & MISKITO              &       2 &       1 &       2  \\
243 & GOAJIRO              &       1 &       2 &       2  \\
244 & YABARANA             &       1 &       1 &       2  \\
245 & CHIBCHA              &       3 &       3 &       1  \\
246 & ALACALUF             &       1 &       1 &       3  \\
247 & APINAYE              &       1 &       1 &       2  \\
248 & TUPINAMBA            &       2 &       2 &       2  \\
\hline
\end{tabular}
\end{table}
\newpage
\end{document}